\documentclass[aps,amsmath,amssymb]{revtex4}

\usepackage{graphicx}
\usepackage{bm}
\usepackage[utf8]{inputenc}
\usepackage[T1]{fontenc}

\newcommand\eqn[1]     {Eq.\,(\ref{#1})}

\newcommand\fig[1]     {Fig.\,{\ref{#1}}}
\newcommand\sect[1]    {Sect.\,{\ref{#1}}}

\newcommand{\be}{\begin{equation}}
\newcommand{\ee}{\end{equation}}
\newcommand{\bea}{\begin{eqnarray}}
\newcommand{\eea}{\end{eqnarray}}

\newcommand{\hf}{\frac{1}{2}}

\def\Tr{\mathrm{Tr}}

\newcommand{\fdd}[3]{\frac{\delta^{2} #1}{\delta #2 \delta #3}}

\def\mr#1{{\mathrm{#1}}}

\def\dk{\Delta k}

\def\hp2{\begin{pmatrix}\phi^2_+&0\cr0&\phi^2_-\end{pmatrix}}

\DeclareMathOperator{\arctanh}{arctanh}

\begin{document}

\title{Quantum-classical transition in the Caldeira-Leggett model}
\author{J. Kov\'acs$^{a,b}$, B. Fazekas$^c$, S. Nagy$^a$, K. Sailer$^a$\\
$^a$Department of Theoretical Physics, University of Debrecen, P.O. Box 5, H-4010 Debrecen, Hungary\\
$^b$Institute of Nuclear Research, P.O.Box 51, H-4001 Debrecen, Hungary\\
$^c$ Institute of Mathematics, University of Debrecen, P.O. Box 12, H-4010 Debrecen, Hungary}
\date{\today}

\begin{abstract}
The quantum-classical transition in the Caldeira-Leggett model is investigated in the framework of the functional renormalization group method. It is shown that a divergent quadratic term arises in the action due to the heat bath in the model. By removing the divergence with a frequency cutoff we considered the critical behavior of the model. The critical exponents belonging to the susceptibility and the correlation length are determined and their independence of the frequency cutoff and the renormalization scheme is shown.
\end{abstract}

\maketitle

\section{Introduction}

The Caldeira-Leggett (CL) model is a simple semi-empirical model to treat a quantum mechanical anharmonic oscillator which is linearly coupled to a heat bath consisting of a set of noninteracting or independent harmonic oscillators. The model enables us to investigate dissipation phenomena in the framework of quantum mechanics \cite{Caldeira1,Caldeira2}. Although the model is relatively simple, it enables us to reproduce several properties of realistic dissipative systems without running into difficulties that can appear in the usual way of quantization. Furthermore, the interaction-independent construction of the model opens up the possibility of application to a wide range of physical systems exhibiting dissipative phenomena. The heat bath can be considered as the environment of the physical system, which now corresponds to the anharmonic oscillator. As is known the environment plays a crucial role in the appearance of the decoherence phenomenon, the manner by which a quantum system turns effectively  into a classical one. Its coupling to the environment makes the physical system an open quantum system in which quantum dissipation and entanglement may occur, so that the CL model may provide a good framework to discuss such phenomena, too.

The microscopic interaction between the oscillator and the heat bath is quadratic in the frequency therefore it can be integrated out, giving an additional non-local term into the action. The resulting effective theory can be investigated by several techniques. Here we apply the functional renormalization group (RG) method which enables one to eliminate the degrees of freedom systematically \cite{Wetterich,Morris,BTW,Delamotte,Polonyi,Pawl,Gies,Nagy_rg}. The method starts from a high-energy ultraviolet (UV) microscopic action, and gives its evolution into the infrared (IR) region. Thus the method can treat how the physical system changes throughout several orders of magnitude of the energy scale into the low energy regime. Furthermore, the RG method is very powerful in treating phase transitions, therefore we expect that this method can give a proper description for the quantum decoherence in the CL model \cite{Aoki1,Aoki2,Aoki3}.

The quantum mechanical anharmonic oscillator can be considered as the $d=1$ dimensional version of the $O(1)$ model in quantum field theory. Generally the $O(N)$ model has a Wilson-Fisher (WF) fixed point which separates two phases, the symmetric and the spontaneously broken (or simply broken) ones. In the case of $d=1$ the model has a single phase, since there is no spontaneous symmetry breaking due to the quantum tunneling effect \cite{Kapoyannis,Zappala,Nagy_qmosc,Kovacs_css,Marian,Mati}. However, the RG treatment runs into difficulty when the initial double-well potential meets a weak anharmonic coupling \cite{Zappala}. The numerical solution of the Schr\"odinger equation for the anharmonic oscillator shows up a convex effective potential, independently of the strength of the anharmonic coupling. In the case of the RG method we should use higher orders of the gradient expansion in order to get acceptable results, nevertheless there always remains a region of the parameter space where the effective potential is concave, independently of the renormalization scheme \cite{Kovacs_css,Marian}.

However, in the CL model the heat bath makes the oscillator to an open quantum system, where the spontaneous breakdown can take place. The heat bath can dissipate the quantum fluctuations, which would suppress the quantum tunneling, therefore the model can have a concave effective potential. This makes the RG method in the CL model more effective than usual even in the lowest-order approximation of the gradient expansion, i.e. in the local potential approximation (LPA). We note that the calculations by the instanton method also confirm the existence of this phase in the CL model \cite{Aoki2}.

The dissipation of the quantum fluctuations has already been investigated in the literature \cite{Aoki1}. Since the heat bath is quadratic in the field variable it can be integrated out analytically, which gives a new term into the action, however the frequency integral for the modes, the spectral function, in the heat bath leads to a UV divergence. It is common to identify its local part which can be eliminated by redefining the mass parameter of the anharmonic oscillator. In the case of the Callan-Symanzik RG scheme it may happen, however that the regulator does not eliminate the UV divergencies \cite{Nagy_sg}. Although the results can be brought into accordance with the ones that are obtained by a proper regulator, we usually avoid the mixing of the functional renormalization with the additive one. This motivates us to exchange the additive renormalization by a simple cutoff in the frequency integration in order to remove the UV divergences. We use either a simple cutoff in the form of a unitstep function, or a Lorentzian function, both containing a frequency cutoff of the spectrum of the heat bath. According to physical arguments this frequency cutoff is small and far from the UV cutoff of the renormalization.

Our goal is to determine some critical exponents of the CL model by using various types of the frequency cutoff in the spectral function as well as various renormalization methods and investigate their dependence on the frequency cutoff and the renormalization scheme. We invented two functional RG methods: the Wegner-Houghton equation for the blocked action with the gliding sharp cutoff $k$ \cite{WH} and the Wetterich equation which is based on the scale dependence of the effective action \cite{Wetterich,Morris}. In the latter case the Litim regulator \cite{Litim} is used since it provides us an analytic evolution equation. We compare the results that were obtained by either the WH or the Wetterich formalisms, respectively.

In \sect{sect:CL} we briefly present the Caldeira-Leggett model. We shortly introduce the Wegner-Houghton and the Wetterich renormalization group equations in \sect{sect:RG}. The results are summarized in \sect{sect:res} and the conclusions drawn up in \sect{sect:conc}.

\section{Construction of the Caldeira-Leggett model}\label{sect:CL}

The CL model is built up as follows. The physical system is a simple quantum mechanical anharmonic oscillator and its environment is simulated by an infinite system of harmonic oscillators which constitutes the heat bath, in Minkowski spacetime the action is
\bea
S=\int\left(\hf M{\dot q}^2-V(q)+\sum_n\frac{1}{2}m_n{\dot q_n}^2-\sum_n\frac{1}{2}m_n\omega_n^2q_n^2+q\sum_nC_nq_n\right).
\eea
Here $q$ with no subscript stands for the field variable of the anharmonic oscillator, $V(q)$ is the corresponding potential, $M$ is its mass. The environmental modes are denoted by $q_n$ (with mass $m_n$ and frequencies $\omega_n$) and the coupling between the system and the environment is governed by the coupling $C_n$. The spectral function 
\bea
J(\omega)=\sum_n \frac{C_n^2}{4m_n\omega_n}2 \pi \delta(\omega-\omega_n)
\eea 
is introduced in order to describe the frequency dependence of the individual oscillators in the heat bath. Instead of the discrete spectrum we introduce a continuous one describing the Ohmic dissipation via
\bea\label{osf}
J_\Omega(\omega)=\eta\omega,
\eea
which has a linear form in the frequency. It leads to the systems's equation of motion
\bea\label{eom}
M\ddot{q}=-V'(q)-\eta \dot q,
\eea
where the usual classical dissipation term can be identified, with the positive damping constant $\eta$. We note that in this case there is no cutoff in the spectrum, therefore  one should introduce a proper counterterm into the equation of motion in \eqn{eom} \cite{Aoki1}. Instead, we introduce a frequency cutoff into the spectral function $J(\omega)$. We choose
\bea\label{Junit}
J_u(\omega)= \eta\omega\theta(\Lambda_u-\omega),
\eea
where a unitstep function plays the role of the cutoff. We also choose another form of the cutoff according to
\bea\label{Jlorentz}
J_l(\omega)= \eta\omega\frac{\Lambda_l^2}{\Lambda_l^2+\omega^2}.
\eea
Here the Lorentzian function has a local maximum, although it covers the whole frequency interval. The unitstep- and the Lorentzian type distributions can be identified by the Debye- and Drude type distributions in the phonon spectrum, respectively. Naturally the simple classical dissipation form as was got in \eqn{eom} changes. Following \cite{Aoki1} we investigate the quantized model in Euclidean spacetime. The heat bath contributes to the action with the term
\bea\label{DeltaS}
\Delta S=-\hf \sum_n \int \bar q(\omega) \frac{C_n^2}{m_n (\omega^2+\omega_n^2)}\bar q(-\omega) d\omega= \hf \int \bar q(\omega) \Sigma(\omega)\bar q(-\omega) d\omega,
\eea
where $\bar q$ is the Fourier transform of the system coordinate $q$.

In \eqn{DeltaS} the self-energy
\bea
\Sigma(\omega)=-\frac{1}{2\pi}\int_0^{\infty}d\omega' J(\omega')\frac{4\omega'}{\omega^2+\omega'^2}
\eea
has also been introduced. We should perform the integration over the frequency by using different forms of the spectral functions. Since we shall use the RG technique we exchange its frequency variable $\omega$ into the renormalization scale $k$. For the unitstep cutoff we obtain that
\be\label{usf}
\Sigma_u(k)=-\frac{2\eta}{{\pi}}\bigg(\Lambda_u-k\arctan\frac{\Lambda_u}{k}\bigg).
\ee
For the Lorentzian function the self-energy becomes
\be\label{lsf}
\Sigma_l(k)=-\frac{\eta\Lambda_l^2}{\Lambda_l+k}.
\ee
The value of $\eta$ controls the strength of the interaction between the anharmonic oscillator and the heat bath. For weak interactions we expect that there is no broken phase in the model. We do not know in advance whether the critical value of $\eta_c$, where the broken phase appears is small or large, therefore we cannot treat the problem perturbatively in $\eta$. We do not consider $\eta$ as a scale-dependent coupling, instead we consider it as a simple constant. In the RG treatment the evolution of the couplings will depend on $\eta$. 

\section{Functional renormalization group equations}\label{sect:RG}
\subsection{Wegner-Houghton equation}\label{sect:WHRG}

We use functional RG methods in order to determine the critical exponents of the CL model. In this section we give a short review of the WH equation. It is based on the evolution of the blocked action. We take the loop expansion of the functional integral and then we perform a general blocking step which integrates out the modes that are characterized by the scale $k$ to $k-\dk$. Starting from the Euclidean Wilsonian action $S_k[q]$ the blocking step results in the following form
\be\label{blrel}
e^{-S_{k-\dk}[q]}=e^{-S_k[q+q']-\hf\Tr\ln\fdd{S_k[q+q']}{q}{q}},
\ee
where the field variable $q'$ stands for the saddle point. The initial condition is imposed at the cutoff $k=\Lambda$. The trace takes into account the modes within the momentum shell $p\in [k,k-\dk]$
\be
\label{tr}
\Tr\ln\fdd{S_k[q+q']}{q}{q}=\int_{k-\dk<|p|<k}\ln\fdd{S_k[q+q']}{q_{-p}}{q_{p}}.
\ee
We assume that there is a trivial saddle point field configuration $q'=0$. In the local potential approximation (LPA) the action of the model consists of the kinetic term and the local potential $U_k$ in the form of
\be
S_k[\phi]=\hf\int_x M{\dot q}^2+\int_xV_k(q)+\hf\int_x \Sigma_k q^2.
\ee
In accordance with the LPA we have set $M=1$. Fortunately the integration of the modes for the momentum shell can usually be performed analytically. This may give a great advantage compared to the Wetterich equation. After performing the momentum-shell integral we arrive at the WH equation
\be
\label{WHeucl}
\dot V_k = -\frac{k}{2\pi} \ln\left(k^2+V''_k+\Sigma_k\right),
\ee
where the dot is a short notation of $k\partial_k$, furthermore $V''_k =\partial^2_q V_k$. In the LPA ansatz
\bea\label{pot}
V_k= \hf m_k^2 q^2+\frac{1}{4!}g_k q^4+\sum _{n=3}^N \frac {g_{2n}(k)}{(2n)!} q^{2n}
\eea
for the potential we take into account further couplings beyond the quartic term (up to $N=6$) in order to achieve a proper convergence in the evolution of the couplings $m_k^2$ and $g_k$.

\subsection{Wetterich equation}

Wetterich's approach describes the evolution of the effective average action $\Gamma$ \cite{Wetterich,Morris}. The RG equation in Euclidean spacetime is
\bea\label{Wett}
\dot \Gamma_k=\hf\mr{Tr}\frac{\dot R_k}{R_k+\Gamma''_k},
\eea
where the trace Tr signals the integration over all momenta. The IR regulator function $R_k$ stands for removing the UV and IR divergences if necessary. In LPA  we have the evolution equation
\bea\label{WettRG}
\dot V_k=\int_p\frac{\dot R_k}{p^2+V''_k+\Sigma_k+R_k}.
\eea
 for the potential. It is worth using the Litim regulator \cite{Litim}
\be\label{litreg}
R_k=(k^2-p^2)\theta(k^2-p^2),
\ee
enabling one to perform the momentum integration analytically in certain cases.

\section{Results}\label{sect:res}

\subsection{Wegner-Houghton renormalization scheme}
By using the ansatz \eqn{pot} for the potential, we derive a system of differential equations for the couplings, which should be solved numerically.  For the initial conditions, the couplings beyond the quartic one were set to zero. The flow equations for the couplings contain the self-energy $\Sigma_k$ which has been chosen in various forms. First we use the one which corresponds to the spectral function that describes the ohmic dissipation without frequency cutoff, i.e.
\be
\Sigma_k = \eta k.
\ee
We take various  fixed values for $\eta$ so that the flow of the couplings should depend on it. The coupling $\eta$ has the dimension of the mass. If the model has a single symmetric phase, then it implies that independently of the initial conditions the inverse propagator remains positive during the evolution. Generally the inverse propagator is
\be
G^{-1}_k = k^2+{\cal M}_k^2,
\ee
where the dressed mass ${\cal M}_k^2$ contains the mass coupling $m_k^2$ and the frequency cutoff dependent contribution from the self energy. In the deep IR regime the dressed mass should tend to positive values in the symmetric phase. Naturally, when the initial mass is negative and the quartic coupling is small, then one might face the problem that appeared in the RG analysis of the anharmonic oscillator \cite{Nagy_qmosc}.  For small values of $\eta$ it is expected that we have a symmetric phase. There is a critical value $\eta_c$ of $\eta$, where the broken phase appears. We can identify the broken phase from the fact that the inverse propagator becomes zero.

\begin{figure}[ht]
\begin{center}
\includegraphics[width=8cm,angle=-90]{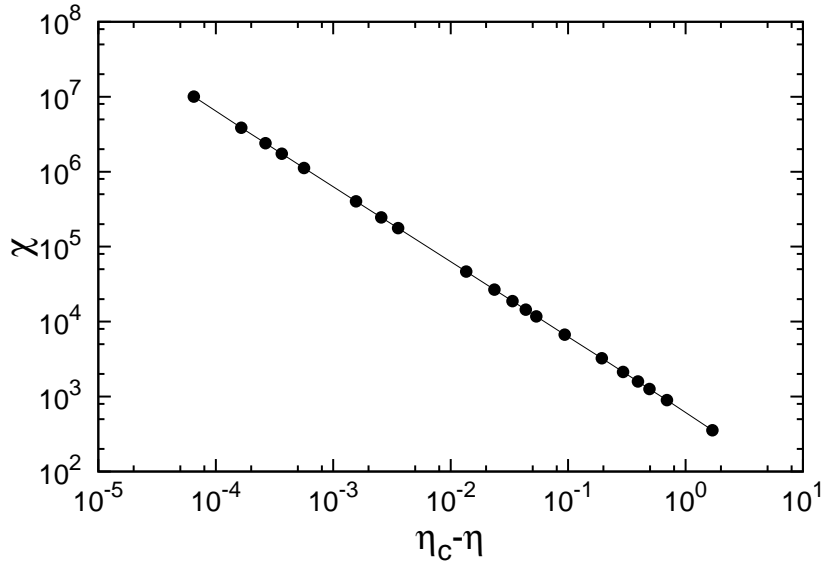}
\caption{\label{fig:whogam} The scaling of the susceptibility $\chi$ is shown near the critical value $\eta_c$. In log-log scale the slope of the curve gives the value $\gamma=1$ of the critical exponent.}
\end{center}
\end{figure}

We calculated the exponent $\gamma$ of the susceptibility, which is defined by
\be
\chi = G_{k\to 0} = \frac1{{\cal M}^2_0},
\ee
so we should determine the mass in the IR limit. In the vicinity of the transition point the mass approaches smaller and smaller positive values, and the susceptibility starts to diverge. The exponent is defined as the degree of divergence, i.e.
\bea
\chi \sim |\eta-\eta_c|^{-\gamma},
\eea
where $\eta_c$ separates the phases. Above $\eta_c$ the model has a broken phase. If the coupling between the oscillator and the heat bath is strong enough, then quantum tunneling cannot appear. The strong coupling enables the environment to dissipate the quantum fluctuations, and the RG analysis describes a quantum-classical transition, where the quantum tunneling effects do not occur, and a simple discrete $Z_2$ symmetry breaking takes place. Let us notice that the critical point is approached from the side of the symmetric regime. The plot in \fig{fig:whogam} shows how the mass diverges as $\eta$ approaches  $\eta_c$ from below.
In the log-log scale we have read off the value of the exponent $\gamma$ and obtained the value $\gamma=1$ with high accuracy. This result deviates from the one obtained in \cite{Aoki1}, however we have used a different technique in order to identify the exponent.

Also the exponent of the correlation length has been determined numerically. Usually the correlation length $\xi$ is identified as the reciprocal of the scale $k=k_c$ where the evolution stops in the broken phase \cite{Nagy_ir,Braun}. It is obvious that the critical behavior is now detected when the transition point is approached from the side of the broken phase, as is usual. There the correlation length is finite and starts to diverge as $\eta\to\eta_c$ according to
\bea
\xi \sim |\eta-\eta_c|^{-\nu}.
\eea 
In the log-log plot  of $\xi$ as the function of $\eta-\eta_c$ we can see the critical behavior similar to the one shown in \fig{fig:whogam}. In that way the value $\nu=1$ have been obtained.

\begin{figure}[ht]
\begin{center}
\includegraphics[width=8cm,angle=-90]{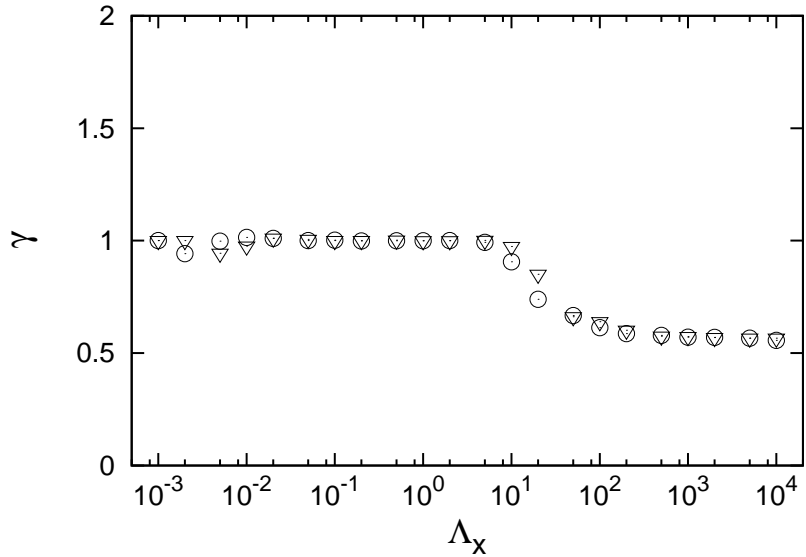}
\caption{\label{fig:gamwh} The value of the critical exponent $\gamma$ as the function of the frequency cutoff is shown. The subscript $x$ stands for $u$ and $l$. The triangles (circles) show the values of the exponents for the unitstep (Lorentzian) cutoff, respectively. We set the initial conditions as $m^2(\Lambda)=-1$, $g(\Lambda)=1$, $\Lambda=10^4$.
}
\end{center}
\end{figure}

Let us now  turn to  the cases when we choose spectral functions  containing a cutoff in the frequency spectrum of the heat bath. We calculated the same exponents as in the previous case. Besides the $\eta$-dependence, the exponents depend on the frequency cutoff, as well. The self-energy coming from the spectral function incorporating the unitstep cutoff function is basically a restricted version of the ohmic dissipation. The exponent $\gamma$ of the susceptibility is plotted in \fig{fig:gamwh} vs. a  broad range of the cutoff $\Lambda_u$.
When $\Lambda_u$ is small we have $\gamma=1$ for the exponent, so the IR limit of $\Lambda_u$ can give back the result obtained in the ohmic dissipation. In the UV limit of $\Lambda_u$ the exponent tends to another value, namely $\gamma\approx 0.57$. We showed numerically that $\Lambda_x$ and $\eta_c$ are strongly correlated, they exhibit a power law relation according to $\eta_c\sim\Lambda_x^{-1}$. The region where the frequency cutoff is comparable to the scale of $\eta_c$ coincides with the transition region, where the value of the exponent $\gamma$ in \fig{fig:gamwh} changes rapidly. In the UV limit the renormalization and the frequency cutoffs are not separated well, therefore this limit has no physical relevance, and the IR regime of the frequency cutoff can be considered as the physical region. Fortunately there is a broad interval across several orders of magnitude, where the value of the exponent is $\gamma=1$, and independent of $\Lambda_u$. As \fig{fig:gamwh} shows, we obtain the same results when we use the Lorentzian cutoff. We can conclude that the exponent $\gamma$ can be calculated independently of both the value of the frequency cutoff and the manner the spectrum of the heat bath is cut off, and  numerics yield the unique value $\gamma=1$. 

Let us now turn to the calculation of the critical exponent of the correlation length. As we approach the IR limit, two types of evolution may arise. When the initial conditions give flows belonging to the symmetric phase, then we can reach the limit $k\to 0$. Those flows which correspond to the broken phase run into singularities. In the broken phase the dressed mass is negative, therefore the inverse propagator can be zero. The propagator and its powers appear in the $\beta$ functions of the couplings, thus the flows diverge. Theoretically the RG flow cannot have singularities, however the truncation of the model in the gradient expansion (we use LPA) and the truncated Taylor expansion of the potential cause singularities. It manifests itself in the fact that there occurs a critical scale $k_c$, where the evolution stops \cite{Pangon,Nagy_rg,Braun}. The stopping scale $k_c$ represents a characteristic momentum in the theory, since there a bulk amount of zero-energy excitations appears. These modes build up a condensate, therefore it is plausible to assume, that the scale $k_c$ characterizes the reciprocal of the size of the condensate itself.  We identify the correlation length as the reciprocal of the stopping scale, i.e. $\xi\sim 1/k_c$.

\begin{figure}[ht]
\begin{center}
\includegraphics[width=8cm,angle=-90]{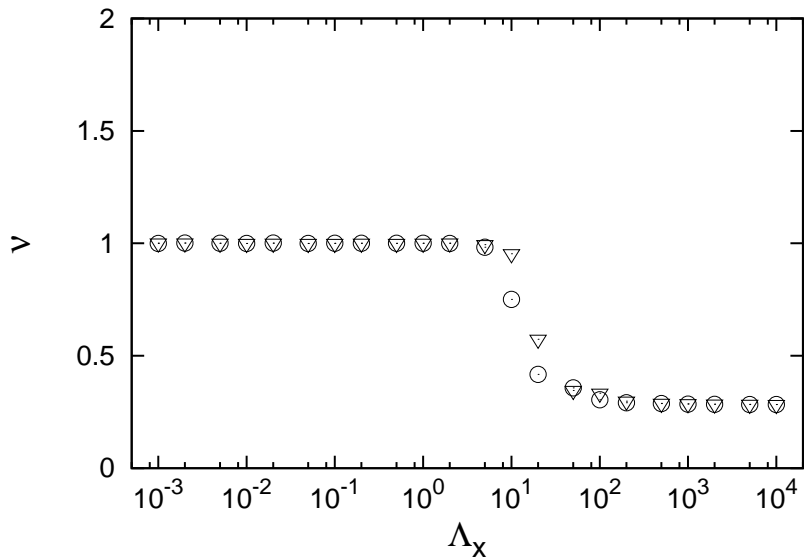}
\caption{\label{fig:nuwh} We present the value of the critical exponent $\nu$ for different values of the frequency cutoff. The subscript $x$ stands for $u$ and $l$. The circles (triangles) show the values of the exponents for the unitstep (Lorentzian) cutoff, respectively. We set the initial conditions as $m^2(\Lambda)=-1$, $g(\Lambda)=1$, $\Lambda=10^4$.
}
\end{center}
\end{figure}

We calculated the stopping scale for different values of the frequency cutoff and determined the critical exponent $\nu$. The results are plotted in \fig{fig:nuwh}.
It can be seen that the unitstep- and the Lorentzian cutoffs give the same results. In the IR limit the critical exponent tends to $\nu=1$. There are several orders in the magnitude of $\Lambda_u$, where the exponent keeps its  constant value. Similarly to the values of $\gamma$ we have an intermediate region around $\Lambda_u\approx\eta_c$, where there is a strong cutoff dependence. In the UV limit we have $\nu\approx 0.3$, but the IR limiting value is physically relevant, again.

We can conclude that by using the WH equation the critical exponents $\gamma$ and $\nu$ can be calculated consistently. Their values do not depend either on the value of the frequency cutoff or the type of the cutoff function.

\subsection{Litim regulator}

\begin{figure}[ht]
\begin{center}
\includegraphics[width=8cm,angle=-90]{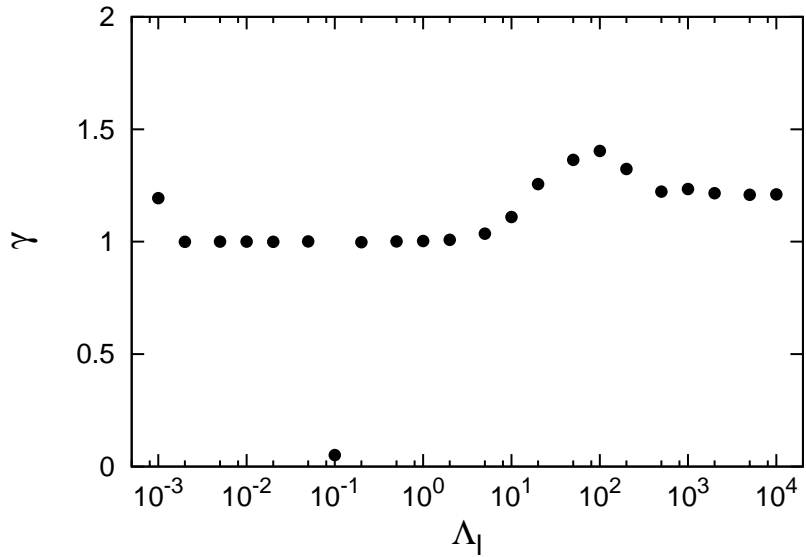}
\caption{\label{fig:gamlit} The value of the critical exponent $\gamma$ is plotted for the Litim regulator. We set the initial conditions as $m^2(\Lambda)=-1$, $g(\Lambda)=1$, $\Lambda=10^4$.
}
\end{center}
\end{figure}

In order to check the scheme independence of the exponents we derived the evolution equations in the framework of the Wetterich equation. We start from the general LPA evolution equation of the potential in \eqn{WettRG} and we use the Litim regulator in \eqn{litreg}. For the self-energy we first choose the ohmic dissipation in \eqn{osf}. Fortunately the loop integral in the Wetterich equation can be performed analytically and we find
\bea
\dot V_k=\frac{k^2}{\eta\pi}\ln\bigg(1+\frac{\eta k}{k^2+V''_k}\bigg).
\eea
In the $\eta\to 0$ limit we can recover the usual form of the Wetterich RG equation with the Litim regulator. We concentrated on the calculation of the exponent $\gamma$. We obtained that $\gamma=1$ similarly to the WH scheme results. Then we determined the value of $\gamma$ for the self-energy containing the Lorentzian like spectral function corresponding to  \eqn{lsf}. We derived the evolution equation
\bea
\dot V_k=\frac{k^2}{\pi(k^2+V''_k)^2}\left(k^3+2\eta\Lambda_l^2\arctanh\frac{k(k^2+V''_k)}
{k^3+2k^2\Lambda_l-2\eta\Lambda_l^2+(k+2\Lambda_l)V''}+kV''_k\right).
\eea
The dependence of the critical exponent $\gamma$ on the frequency cutoff can be seen in \fig{fig:gamlit}.
In the IR limit of $\Lambda_l$ we obtained the stable value $\gamma=1$ in good agreement with the WH scheme results. Beyond the transition region we obtain $\gamma\approx 1.2$ in the UV limit, which deviates from the previous results. We note that the UV value is not universal and this signals that frequency-cutoffs $\Lambda_l>\eta_c$ are not physical, as discussed above. The determination of the exponent $\nu$ runs into numerical difficulties since one should consider the calculation in the broken phase. Nevertheless the correct IR value for the exponent affirms that $\gamma=1$ indepently on the RG scheme and the spectral function.

\section{Conclusions}\label{sect:conc}

We investigated the Caldeira-Leggett model in the framework of various functional renormalization group methods. Critical exponents have been studied near the quantum-classical phase transition, where the heat bath dissipates the quantum effects.

The critical exponent $\gamma$ of the susceptibility has been determined for various types of the spectral function describing the frequency dependence of the modes in the heat bath with and without a frequency cutoff. Furthermore we derived the evolution equation for the local potential in the framework of the Wegner-Houghton RG scheme as well as  in that of  Wetterich's  effective average action method. In the latter case we succeeded in deriving a closed form of the evolution equation by using the Litim regulator. The universal value $\gamma=1$ has been obtained  in every case independently of the RG scheme, of either neglecting or accounting for the frequency cutoff of the spectrum of the environment, and of the manner the latter is taken into account.

We also determined the value of the exponent $\nu$ belonging to the correlation length. We used the Wegner-Houghton equation and we got the value $\nu=1$ independently of the various choices of the spectral function.

Our investigation showed that the proper treatment of the spectral function with a physical frequency-cutoff seemingly introduces the dependence of the critical exponents on that frequency-cutoff, but it turns out that the numerically determined critical exponents $\gamma$ and $\nu$ become independent of the cutoff of the spectrum of the heat bath when the latter  is much lower than the cutoff $\Lambda $ of the bare theory. Then the values of $\gamma$ and $\nu$ are universal. 

\section*{Acknowledgments}
S. Nagy acknowledges financial support from a J\'anos Bolyai Grant of the Hungarian Academy of Sciences, the Hungarian National Research Fund OTKA (K112233).

\end{document}